\documentclass[12pt]{article}%
\usepackage{amsmath}
\usepackage{amsfonts}
\usepackage{amssymb}
\newtheorem{thm}{Theorem}

\newenvironment{proof}[1][Proof]{\noindent \textbf{#1.} }{\  \rule{0.5em}{0.5em}}
\begin{document}

\date{}

\title{Late time behavior of closed isotropic models in second order
gravity theory}

\author{John Miritzis,\\Department of Marine Sciences, University of the Aegean,\\University Hill, Mytilene 81100, Greece}
\maketitle

\begin{abstract}
Homogeneous and isotropic closed models are studied in both the Einstein and the
Jordan frame of the second order gravity theory. The normal form of the dynamical
system has periodic solutions for a large set of initial conditions. This implies that
an initially expanding closed isotropic universe may exhibit oscillatory behavior.
\end{abstract}


In this paper we investigate the late time evolution of positively curved FRW
models with a perfect fluid in the $R+\beta R^{2}$ theory. The simple vacuum
case was studied in \cite{miri1}, where oscillatory behavior of the solutions
of closed models was found. Since there is no universally acceptable answer to
the issue \textquotedblleft which conformal frame is
physical\textquotedblright, (see for example \cite{maso}), it should be
interesting to compare the solution in the Jordan frame with the solution of
the same problem obtained in the Einstein frame. In the following, we confine
our attention to cosmologies with a perfect fluid with energy density $\rho$
and pressure $p,$ of the form $p=(\gamma-1)\rho,\ \ \ \ 0\leq\gamma\leq2.$

We begin our study in the Einstein frame. The scalar field is minimally
coupled to ordinary matter. The Einstein equations reduce to the constraint
\[
3H^{2}+\frac{3k}{a^{2}}=\rho+\frac{1}{2}\dot{\phi}^{2}+V\left(  \phi\right)
,
\]
and the evolution equations%
\begin{gather*}
\dot{a}=aH,\ \ \ \dot{H}=-\frac{1}{3}\dot{\phi}^{2}-\frac{\gamma}{2}\rho
+\frac{k}{a^{2}},\ \\
\dot{\rho}=-3\gamma\rho H,\ \ \ \ddot{\phi}+3\frac{\dot{a}}{a}\dot{\phi
}+V^{\prime}\left(  \phi\right)  =0,
\end{gather*}
where the potential takes the form%
\[
V\left(  \phi\right)  =\frac{1}{8\beta}\left(  1-e^{-\sqrt{2/3}\phi}\right)
^{2}.
\]
We use the constraint to eliminate $a,$ rescale the variables and set
$\dot{\phi}=:y,\ \ \ u:=e^{-\phi},$ to obtain the quadratic system%
\begin{align*}
\dot{u} &  =-uy,\ \ \ \dot{y}=-u+u^{2}-3Hy,\\
\dot{\rho} &  =-3\gamma\rho H,\ \ \ \ \ \ \dot{H}=\frac{1}{4}\left(
1-u\right)  ^{2}-\frac{1}{2}y^{2}-\frac{3\gamma-2}{6}\rho-H^{2}.
\end{align*}

There are two equilibria, $\mathcal{S}$,\ corresponding to the de Sitter
universe with a cosmological constant equal to $\sqrt{\frac{1}{8\beta}}$ and
$\mathcal{M}$,$\ $ corresponding to the limiting state of an ever-expanding
universe with $H\rightarrow0$ while the scalar field approaches the minimum of
the potential and the scale factor goes to infinity.

\begin{thm}
The equilibrium point $\mathcal{S}$ is locally asymptotically unstable.
\end{thm}

\begin{proof}[Sketch of the proof]
We apply the center manifold theorem, according to which the qualitative behavior in a
neighborhood of a nonhyperbolic equilibrium point $\mathbf{q}$ is determined by its
behavior on the center manifold near $\mathbf{q.}$ We compute the center manifold and
find the restriction of the dynamical system on it; it turns out that $\mathcal{S}$ is
unstable.
\end{proof}

$\mathcal{M}$ is totally degenerate. Applying the normal form theory, one may
show that an initially expanding closed universe near $\mathcal{M}$ cannot
avoid recollapse for $\frac{2}{3}<\gamma\leq2$, see \cite{miri} for details.

Next we return to the Jordan frame. With $x=1/a,$ the $0-0$ equation becomes
\[
H^{2}+kx^{2}+2\beta\left[  R\left(  H^{2}+kx^{2}\right)  +H\dot{R}-\frac
{R^{2}}{12}\right]  =\frac{1}{3}\rho,
\]
and the evolution (fourth-order) equations are
\begin{align*}
\dot{x}  &  =-xH,\ \ \ \ \dot{H}=\frac{1}{6}R-2H^{2}-kx^{2},\\
\dot{\rho}  &  =-3\gamma\rho H,\ \ \ddot{R}+3H\dot{R}+\frac{1}{6\beta}%
R=\frac{1}{6\beta}\left(  4-3\gamma\right)  \rho.
\end{align*}
The state of the system is $(R,\dot{R},x,\rho,H)\in\mathbb{R}^{5}$. We use the
constraint to eliminate $\rho$, so we obtain a four-dimensional system.

There are two equilibria: $\mathcal{S}$ corresponding to the Einstein static
universe, where the effective cosmological constant is provided by the
curvature equilibrium
\[
k\bar{x}^{2}+\frac{\Lambda}{3}=\frac{1}{3}\bar{\rho},\ \ \Lambda>0.
\]
$\mathcal{S}$ is unstable (the local stable and unstable manifolds through
$\mathcal{S}$ are both two-dimensional).

The other equilibrium point is the origin $\left(  0,0,0,0\right)  $.
Linearization shows that the eigenvalues of the Jacobian matrix at the origin
are $\pm i,0,0.$ The normal form of the system near $\left(  0,0,0,0\right)  $
in cylindrical coordinates is%

\begin{align}
\dot{r} &  =-\frac{3}{2}ry,\ \ \ \ \dot{\theta}=1,\label{cyli}\\
\dot{x} &  =-xy,\ \ \ \ \dot{y}=6\left(  \gamma-1\right)  r^{2}-\frac
{3\gamma-2}{2}x^{2}-\frac{3\gamma}{2}y^{2}.\nonumber
\end{align}
We write the first and third equations as a differential equation%
\[
\frac{dr}{dx}=\frac{3}{2}\frac{r}{x}\ \ \Rightarrow\ r=Ax^{3/2},\ \ A>0.
\]
Substitution into the fourth equation of (\ref{cyli}) yields the projection of
the fourth-dimensional system on the $x-y$ plane, namely%
\[
\dot{x}=-xy,\ \ \ \ \ \ \dot{y}=b\left(  \gamma-1\right)  x^{3}-\frac
{3\gamma-2}{2}x^{2}-\frac{3\gamma}{2}y^{2},\ \ \ b>0.
\]
We note that: Trajectories are symmetric with respect to the $x$ axis, the
line $x=0$ is invariant, and the system has a first integral,
\[
\phi\left(  x,y\right)  =-\frac{2b}{3}x^{3\left(  1-\gamma\right)
}+x^{2-3\gamma}+\frac{y^{2}}{x^{3\gamma}},\ \ \gamma\neq1,\ \ \ \ \ \ \phi
\left(  x,y\right)  =\frac{1}{x}+\frac{y^{2}}{x^{3}},\ \ \gamma=1.
\]
The level curves of $\phi$ are the trajectories of the system. In \cite{miri2}
is proved the following result.

\begin{thm}
For every $\gamma\in(\frac{2}{3},2]$ there are no solutions asymptotically
approaching the origin (ii) for $\gamma\in(\frac{2}{3},1]$\ there are no
periodic solutions and (iii) for $\gamma\in(1,2]$ there exist periodic
solutions and the basin of every periodic trajectory is the set
\[
\left\{  \left(  x,y\right)  \in\mathbb{R}^{2}:y^{2}+x^{2}-\frac{2b}{3}%
x^{3}<0, \ \ x>0\right\}  .
\]
\end{thm}

Obviously we cannot assign a physical meaning to the new variables $\left(
r,\theta,x,y\right)$, since the transformations involved in the normal
form have \textquotedblleft mixed\textquotedblright\ the original variables in
a nontrivial way. However, the periodic character of the solutions whatever
the physical meaning of the variables be, has the following interpretation.
\emph{Close to the equilibrium of the original system, there exist periodic
solutions}.

This feature is not revealed in the Einstein frame, possibly because, the
scalar field related to the conformal transformation is not coupled to matter,
i.e., the matter Lagrangian is added after performing the conformal transformation.

In a recent work \cite{migi}, we consider a scalar field having an arbitrary,
bounded from below potential function $V\left(  \phi\right)  ,$ nonminimally
coupled to matter. For flat and negatively curved FRW models, our preliminary
results indicate that the energy exchange between the two matter components is
such that:

\emph{If }$\gamma<1$\emph{\ the energy density }$\rho$\emph{\ eventually
dominates over the energy density of the scalar field }$\epsilon$\emph{\ and
this universe follows the classical Friedmannian evolution. For }$\gamma
>1$\emph{, }$\epsilon$\emph{\ eventually dominates over }$\rho$\emph{.}

\end{document}